\documentclass[aps,twocolumn,showpacs,pra,superscriptaddress,floatfix,longbibliography]{revtex4-1}

\usepackage{amsmath,amssymb,amsmath,amsthm,bm,graphicx,xcolor}
\usepackage{natbib}

\usepackage{color}
\usepackage{tabularx}
\usepackage{epsfig}
\usepackage{amssymb}
\usepackage{graphicx}
\usepackage{wasysym}
\usepackage{dcolumn}
\usepackage{epstopdf}
\usepackage{subfigure}
\usepackage{psfrag}
\usepackage{bbold}
\usepackage[normalem]{ulem}

\usepackage{array}   
\newcolumntype{L}{>{$}l<{$}}

\newcommand{\R}{{\mathbb{R}}}

\newcommand{\bra}[1]{\mbox{$\langle #1 |$}}

\newcommand{\ket}[1]{\mbox{$| #1 \rangle$}}

\begin{document}

\title{Generalized Pauli constraints in small atoms}

\author{Christian Schilling}
\email{Christian.Schilling@physics.ox.ac.uk}
\affiliation{Clarendon Laboratory, University of Oxford, Parks Road, Oxford OX1 3PU, United Kingdom}

\author{Murat Altunbulak}
\affiliation{Department of Mathematics, Faculty of Science, Dokuz Eylul University, 35390 Buca-Izmir, Turkey}

\author{Stefan Knecht}
\affiliation{ETH Z\"urich, Laboratorium f{\"ur} Physikalische Chemie, Vladimir-Prelog-Weg 2, 8093 Z\"urich, Switzerland}

\author{Alexandre Lopes}
\affiliation{Carl Zeiss SMT GmbH, Rudolf-Eber-Stra{\ss}e 2, 73447 Oberkochen, Germany}

\author{James D. Whitfield}
\affiliation{Department of Physics and Astronomy, Dartmouth College, 6127 Wilder Laboratory, Hanover, NH 03755, USA}

\author{Matthias Christandl}
\affiliation{QMATH, Department of Mathematical Sciences, University of Copenhagen, Universitetsparken 5, 2100 Copenhagen, Denmark}

\author{David Gross}
\affiliation{Institute  for  Theoretical  Physics,  University  of  Cologne, Z\"ulpicher Stra{\ss}e 77, 50937 Cologne, Germany}

\author{Markus Reiher}
\affiliation{ETH Z\"urich, Laboratorium f{\"ur} Physikalische Chemie, Vladimir-Prelog-Weg 2, 8093 Z\"urich, Switzerland}

\date{\today}

\begin{abstract}
The natural occupation numbers of fermionic systems are subject to non-trivial constraints, which include and extend the original \emph{Pauli principle}.
A recent mathematical breakthrough has clarified their mathematical structure and has opened up the possibility of a systematic analysis.
Early investigations have found evidence that these constraints are exactly saturated in several physically relevant systems; e.g.\ in a certain electronic state of the
Beryllium atom.
It has been suggested that in such cases, the constraints, rather than the details of the Hamiltonian, dictate the system's qualitative behavior.
Here, we revisit this question with state-of-the-art numerical methods for small atoms.
We find that the constraints are, in fact, not exactly saturated, but that they lie much closer to the surface defined by the constraints than the geometry of the problem would suggest.
While the results seem incompatible with the statement that the generalized Pauli constraints drive the behavior of these systems,
they suggest that the qualitatively correct wave-function expansions can in some systems already be obtained on the basis of a limited number of Slater determinants, which is in line with numerical evidence from quantum chemistry.
\end{abstract}



\maketitle
\section{Introduction}
Pauli's \emph{exclusion principle} \cite{Pauli1925} is a well-known physical principle. Its relevance concerns all length scales, from the subatomic (structure of nuclei) up to the astronomic scale (stability of neutron stars). The most prominent example for its relevance, however, is the \emph{Aufbau principle} underlying the periodic table and explaining the structure of atoms.
In the 1970s, it was found that the fermionic exchange symmetry implies further constraints on natural occupation numbers \cite{Borl1972,Rus2}.
Remarkably, it took several decades before their mathematical structure was finally understood and a complete classification could be derived \cite{Kly2,Kly3,Altun}.
To be more specific, these so-called \emph{generalized Pauli constraints} (GPCs) take the form of linear conditions,
\begin{equation}\label{eq:GPC}
  D_j(\vec{\lambda}) \equiv \kappa_j^{(0)} + \sum_{k=1}^d \kappa_j^{(k)} \lambda_k\geq 0\,,
\end{equation}
on the decreasingly-ordered \emph{natural occupation numbers}  $\vec{\lambda}\equiv (\lambda_k)_{k=1}^d$.
These, in turn, are the eigenvalues of the one-particle reduced density matrix $\rho_1\equiv N \mbox{Tr}_{N-1}[\ket{\Psi_N}\!\bra{\Psi_N}]$.
In Eq.\ (\ref{eq:GPC}), the $\kappa_j^{(k)}$ are the coefficients specifying the $j$th linear inequality with the $N$-fermion quantum state $\ket{\Psi_N}\in \wedge^N[\mathcal{H}_1^{(d)}]$,
where the one-particle Hilbert space $\mathcal{H}_1^{(d)}$ has dimension $d$.
For each \emph{setting} $(N,d)$, there are finitely many GPCs, which define a convex polytope $\mathcal{P}\subset \R^d$.
It is a subset of the \emph{Pauli simplex} $\Delta$ defined by the \emph{original Pauli constraints} $1\geq \lambda_1\geq \ldots \geq \lambda_d\geq 0$.

In an ongoing debate, the physical relevance of the GPCs has been explored and discussed  \cite{Kly1,CS2013,BenavLiQuasi,Kly5,Mazz14,CSthesis,MazzOpen,CSQuasipinning,BenavQuasi2,RDMFT,Mazzagain,Alex,CS2015Hubbard,RDMFT2,CBthesis,CS2016a,NewMazziotti,CS2016b,Mazz16,CBcorr,MazzOpen2,FTthesis,walter2013entanglement,kus,tsanov,CS2018pair,CBreview}.
An early numerical result suggested that a specific state of the Be atom (see below) would saturate some of the GPCs \cite{Kly1}.
Based on this observation, the \emph{pinning} phenomenon has been conjectured:
The variational minimization of the energy functional,
from the point of view of the one-particle picture, may settle on the polytope boundary $\partial \mathcal{P}$.
In analogy to the way the string of a pendulum restricts the movement of the attached mass, the GPCs would then constrain the kinematics of the system to a lower-dimensional space, which would shape the qualitative behavior of the system---e.g.\ its response to external perturbations \cite{Kly1,Kly5,CS2015Hubbard,sawicki}.
%
%
%
%

Due to its striking consequences, a central question is whether exact pinning actually occurs in realistic fermionic quantum systems.
While analytic results for a harmonic model system suggest a negative answer \cite{CS2013,CS2016a,CS2016b}, various studies of small atoms and molecules  \cite{Kly1,Mazz14,RDMFT,Mazzagain,CS2015Hubbard,NewMazziotti} seem to confirm the occurrence of pinning in ground states, and hence emphasize the relevance of the GPCs in quantum chemistry.
However, these numerical studies may be inconclusive for two reasons: First, they are based on very small active spaces of only 6 to 10 spin-orbitals and might therefore fail to accurately capture the true physical situation. Second, it has only been recently realized that exact or approximate pinning (\emph{quasipinning}) is in some cases implied merely by the geometry of the polytope $\mathcal{P}$ and the saturation of some of the original Pauli constraints \cite{CSQ}.

In the present work, we revisit the problem and address the above concerns. In particular: (i) We use state-of-the-art numerical methods (involving up to 1000 spin-orbitals) to reproduce the atomic natural occupation numbers to high precision. (ii) We perform a careful analysis of truncation errors. (iii) We employ a precise quantitative measure to distinguish geometrically trivial from non-trivial pinning.

We note in passing that a vast body of numerical data on atoms and molecules obtained in quantum chemistry in the past decades
has led to a classification of electron correlation (and, in turn, of features of sets of natural occupation numbers) that is governed by
the external potential generated by the atomic nuclei in the system. In this context, we recall that the GPCs follow, however, solely from the fermionic exchange symmetry. Just like Pauli's exclusion principle, they are therefore valid for \emph{all} fermionic quantum systems, independent of the concrete physics involved (Hamiltonian).

In order to elucidate in a \textit{quantitative} fashion
the saturation of the GPCs for an electronic structure of the system under investigation,
we restrict ourselves in this work to a study of simple atomic systems such as Li and Be, for which we can provide sufficiently accurate numerical quantum chemical data obtained from full-configuration-interaction (FCI) calculations close to the basis-set limit.

The main result of our analysis is that the states originally investigated in the literature are \emph{not} exactly pinned.
At the same time, the natural occupation numbers lie much closer to the facets of the polytope $\mathcal{P}$ than its geometry
would imply (a phenomenon we referred to as \emph{non-trivial quasipinning} \cite{CSQ}).
This is an important distinction:
Both the original physical interpretation (that the response of pinned systems to perturbations is restricted by the saturated constraints), as well as the most plausible physical mechanism giving rise to pinning in ground states (that the GPCs prevents a further reduction of the ground state energy) require \emph{exact} pinning.
At the same time, recent results \cite{CSQuasipinning, Stability} show that wave functions whose natural occupation numbers lie close to the boundary can be well-approximated by a superposition of a small number of very specific Slater determinants.
This so-called \emph{super-selection rule} for Slater determinants \cite{Kly1} reflects and expands the vast numerical knowledge obtained in the quantum-chemistry community concerning the structure of atomic ground states.
In this sense, generalized Pauli constraints do have physical implications for small atoms, which might, however, be different from
the ones originally anticipated.

\section{Methods}
In order to achieve the results outlined above, we had to meet three technical challenges:

\emph{First}, in order to obtain sufficiently accurate approximations to the natural occupation numbers, we had to find the variational ground states in Hilbert spaces spanned by up to 1000 spin-orbitals (rather than just 6 to 10 as in previous works \cite{Kly1,Mazz14,RDMFT,Mazzagain,CS2015Hubbard,NewMazziotti}).
%
Our numerical optimizations are accurate enough to recover more than 99\% of the correlation energy (As measured for example by a comparison of our best variational energy of \mbox{-14.6667932644 hartree} with the up-to-date best variational energy of \mbox{-14.667356498} hartree for the $^1S$\ spin sector of the Be atom \cite{puch13}).
This consistency check gives us great confidence that, at the very least, the
qualitative results derived from the numerical data (absence of pinning and presence of non-trivial quasipinning) will hold for the exact ground states of the respective atoms as well.
%

\emph{Second}, we had to cope with the problem that the numerics returns up to 1000 occupation numbers (corresponding to the number of spin-orbitals), whereas the GPCs were explicitly known only for settings up to dimension $d=10$.
While systematic algorithms for determining all GPCs exist, the required computational effort increases dramatically with increasing $d$ \cite{Altun,vergne2014inequalities,buergisser2017membership}.
Fortunately, most of the 1000 numerically obtained occupation numbers $\vec\lambda$ are either very close to $1$ or to $0$.
It therefore appears plausible that one can carry out a pinning analysis in terms of a \emph{truncated} version $\vec\lambda'$ of $\vec\lambda$, obtained by ignoring these extreme eigenvalues.
Indeed, a method to obtain quantitative estimates of errors introduced through truncation have recently been developed
\cite{CSthesis,CS2016a}.
To be more precise, we quantify quasipinning by the minimal $l_1$-distance $D_{min}\equiv \mbox{dist}_1\big(\vec{\lambda},\partial \mathcal{P}\big)$ of $\vec{\lambda}$ to the polytope boundary.
We reduce $N$ to $N'$ by ignoring eigenvalues close to $1$, and $d$ to $d'$ by also ignoring those close to $0$.
We denote the minimal distance found in the analysis in the truncated setting $(N',d')$ with polytope $\mathcal{P}'$
by $D'_{min}\equiv \mbox{dist}_1\big(\vec{\lambda}',\partial \mathcal{P}'\big)$.
Under reasonable technical assumptions on the asymptotic behavior of GPCs,
one can relate $D'_{min}$
to $D_{min}$ in the full setting by virtue of the \emph{truncation error} $\varepsilon'$ (see also Appendix \ref{app:truncation}):
\begin{equation}\label{eq:trunc}
\big|D_{min}-D'_{min}\big| \leq \varepsilon'\equiv\sum_{j=1}^{N-N'}(1-\lambda_j)+\!\!\sum_{k=0}^{d-d'-N'+N-1}\!\!\lambda_{d-k}\,.
\end{equation}
Since the minimally required reduced dimension $d'$ allowing for a conclusive pinning analysis turns out to be larger than the maximal $d'$ for which the GPCs were known so far, we have performed extensive calculations to determine the GPCs also for the settings $(N',d')=(3,11),(3,12)$ (see Appendix \ref{app:GPC311}).

\emph{Third}, quasipinning by the \emph{generalized} Pauli constraints can in some cases arise as a mere consequence of quasipinning by the \emph{original} ones.
Indeed, any given set of eigenvalues $\vec\lambda$ is at least as close to $\partial \mathcal{P}$ as it is to $\partial \Delta$
(this holds because all physically attainable boundary points of the Pauli simplex $\Delta$ are also boundary points of $\mathcal{P}$).
It can therefore be argued that pinning to the GPC should only be considered as \emph{non-trivial}, if the distance of $\vec{\lambda}$ to $\partial P$ is much smaller than the distance of $\vec{\lambda}$ to $\partial \Delta$.
To assess the ``degree of non-triviality'' quantitatively, we will employ the so-called \emph{$Q$-parameter} constructed in Ref.~\citenum{CSQ}.

\section{Pinning analysis}
We now explore (quasi)pinning and its non-triviality for several distinct states.

We start with the state which was used as a first example for exact pinning \cite{Kly1}.
In a basis set of five $s$-orbitals for the truncated one-particle Hilbert space, the variational ground state of the Beryllium atom within the
respective spin-triplet sector was calculated in Ref.~\cite{Nakata2001}.
The corresponding natural occupation numbers are listed in Table \ref{tab}\ while the coefficients in the exponents of the atomic-orbital Gaussian basis sets for Be can be found in Table \ref{tabs:basis-sets-be} in Appendix \ref{app:data}).
\renewcommand{\arraystretch}{1.14}
\begin{table}[h]
\begin{tabular}{|L|L|L|L|L|}
\hline
\lambda_1&0.9999999 54956670& &\lambda_6&0.00070683 0007187\\ \hline
\lambda_2&0.9999945 08315023& &\lambda_7&0.00000917 6759273\\ \hline
\lambda_3&0.9992872 03037800& &\lambda_8&0.00000723 7819178\\ \hline
\lambda_4&0.9992838 90609735& &\lambda_9&0.00000010 2559386\\ \hline
\lambda_5&0.0007110 95871330& &\lambda_{10}&0.00000000 0064421\\ \hline
\end{tabular}
\caption{Natural occupation numbers for the variational ground state of the Beryllium atom within the spin-triplet sector.
The data was originally published with six significant digits in Ref.~\cite{Nakata2001}\
but was kindly made available to us in full precision with all 15 significant digits \cite{nakataPrivate}.}
\label{tab}
\end{table}

Based on the data with six significant digits, the pinning analysis could be reduced from the full setting $(N,d)=(4,10)$ to the reduced setting $(3,7)$ apparently
with zero truncation error (cf.\ Eq.~\eqref{eq:trunc}) since $\lambda_1=1.000000$ and $\lambda_9=\lambda_{10}=0.000000$.
Within the reduced setting and by relabeling the indices of the occupation numbers, $\lambda'_1\equiv \lambda_2,\ldots,\lambda'_7\equiv \lambda_8$, the following results for the four GPCs of the reduced setting $(3,7)$ \cite{Borl1972} were found \cite{Kly1}
\begin{eqnarray}\label{eq:GPC37Kly}
       0 &\leq& D_1^{(3,7)} \equiv 2-(\lambda'_1+\lambda'_2+\lambda'_5+\lambda'_6) = 0.000002\nonumber \\
       0 &\leq& D_2^{(3,7)} \equiv 2-(\lambda'_1+\lambda'_3+\lambda'_4+\lambda'_6) = 0.000001\nonumber \\
       0 &\leq& D_3^{(3,7)} \equiv 2-(\lambda'_2+\lambda'_3+\lambda'_4+\lambda'_5) = 0.000011\nonumber \\
       0 &\leq& D_4^{(3,7)} \equiv 2-(\lambda'_1+\lambda'_2+\lambda'_4+\lambda'_7) = 0.000000\,.
\end{eqnarray}
It was striking \cite{Kly1} that summing up the four specific occupation numbers contributing to $D_4^{(3,7)}$
led exactly to the value $2.000000$\ within the numerical precision given.

We now reassess the analysis of the data, using the more precise numerical values for the natural occupation numbers given in Table~\ref{tab}.
It will turn out that the conclusions of the subsequent analysis will require only 8 significant digits.

Exploiting the data in Table~\ref{tab}, we explore pinning in the full setting $(4,10)$, hence incurring no truncation error.
Among the 125 GPCs \cite{Kly3} of that setting, the strongest saturation is found for the constraint
\begin{eqnarray}\label{eq:GPC27}
    0\leq D_{27}^{(4,10)} &\equiv& 4-2 \lambda_1- \lambda_2- \lambda_3- \lambda_5 - \lambda_8\nonumber \\
    & =& 4.5\cdot 10^{-8}\,.
\end{eqnarray}
There are two conclusions to be drawn:
On the one hand, the more precise analysis shows that the seemingly exactly pinned vector $\vec{\lambda}$ does not lie on the boundary of the eigenvalue polytope after all.
On the other hand, it is true that the distance from the boundary is surprisingly small.
Before turning to further examples, we briefly discuss the deviation in Eq.\ (\ref{eq:GPC27}) from exact pinning quantitatively.
For a rough sense of scale, one can compare Eq.~\eqref{eq:GPC27} to the polytope diameter which is on the order of $1$.
However, as explained in the Introduction, we also need to ensure that this degree of quasipinning is non-trivial in the sense of not being implied by the near-saturation of the original Pauli constraints.
For example, according to Table~\ref{tab}, the first four eigenvalues differ from their allowed maximum by no more than $10^{-3}$, which already implies that quasipinning of at least this strength has to be present.
A complete analysis in terms of the $Q$-parameter \cite{CSQ} confirms that the GPCs are, in fact, physically relevant for the electronic state of the Be atom at hand:
The deviation from exact pinning is smaller by a factor of about $245$ from what the value implied merely by the saturation of the Pauli constraints (see Appendix \ref{app:Q}).

The previous result pertained to a variational calculation within a subspace spanned by very few orbitals.
To obtain a more physically accurate description of the Be atom in the triplet state under consideration, one would need to include orbitals
with non-zero angular momentum. 
However, we first pursue a different route by extending only the number of $s$-orbitals with the aim to determine to high precision the variational
ground state within this (artificial) Hilbert space.
The motivation is as follows: Such freezing of degenerate angular degrees of freedom increases the conflict between energy minimization and fermionic exchange symmetry, which has been argued to increase the strength of quasipinning \cite{CS2016a,CS2016b}.
In contrast to the previous analysis of the state involving only five $s$-orbitals the following analyses
will require us to truncate the numerically obtained eigenvalues: It will not be possible anymore to explore (quasi)pinning in the full settings
since the GPCs for four electrons are known only for active spaces spanned by up to five spatial orbitals.

\begin{figure}[]
\centering
\includegraphics[width=4.3cm]{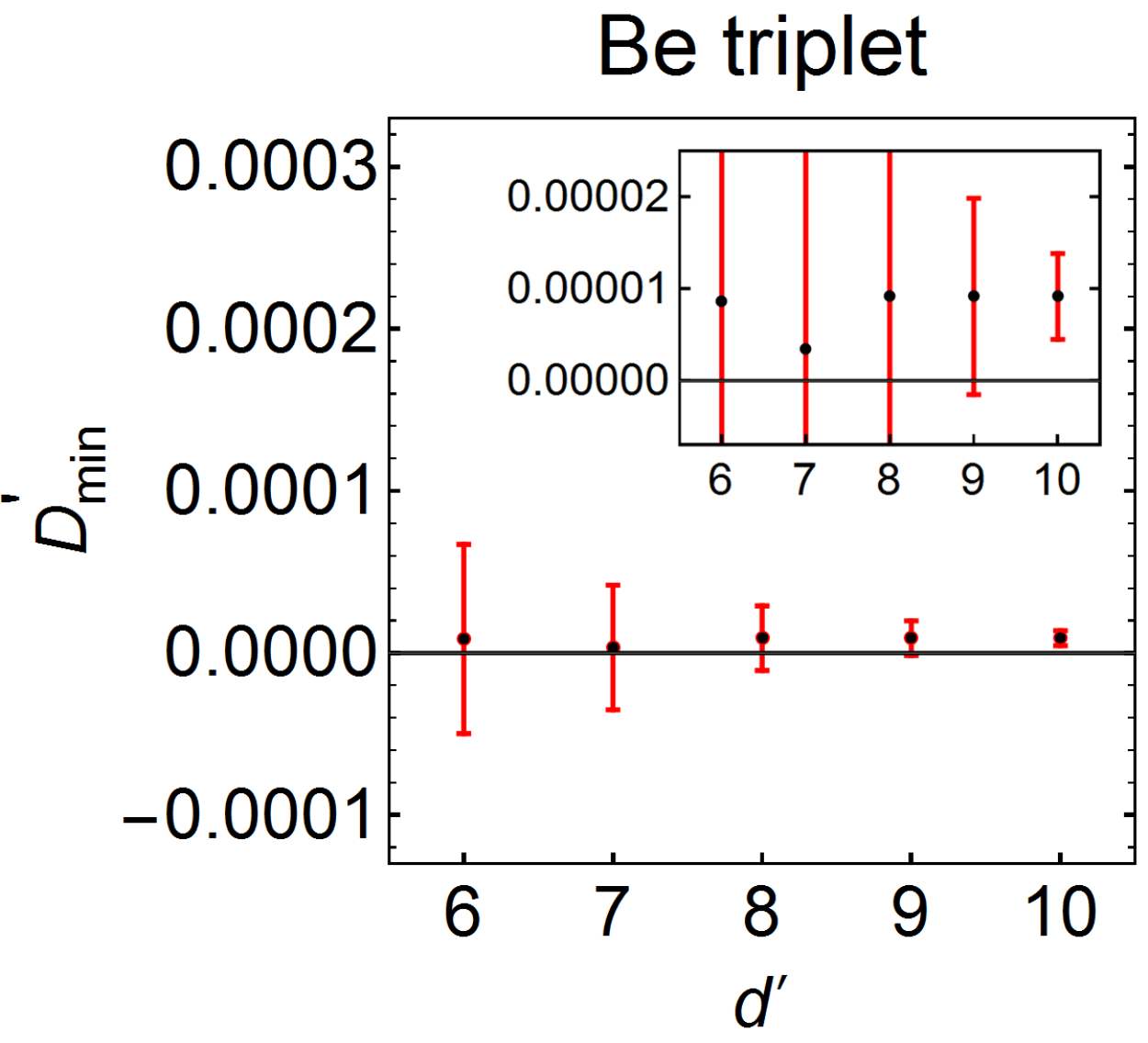}\hspace{0.2cm}
\includegraphics[width=4.0cm]{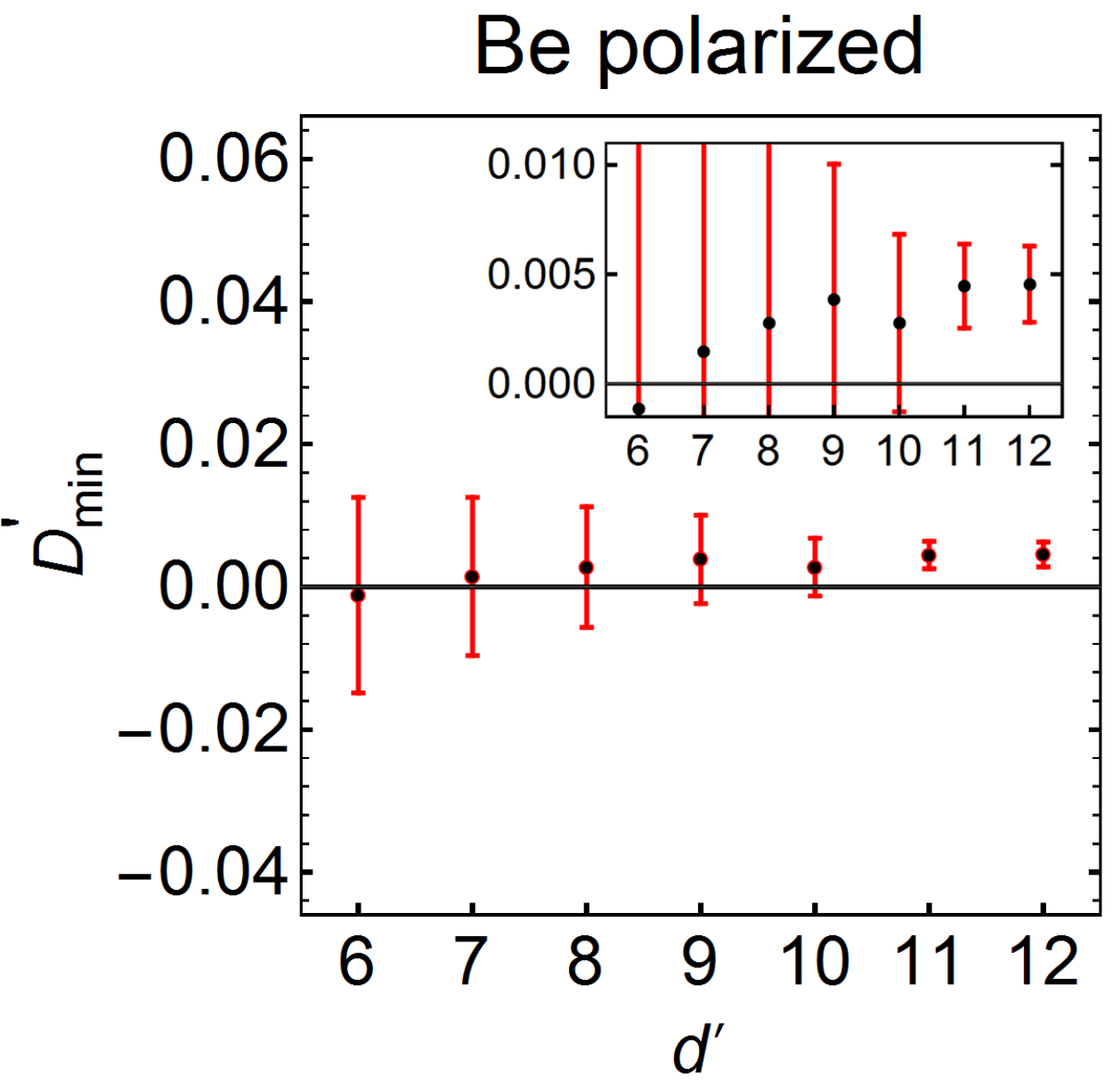}
\caption{Results of the pinning analyses for the variational ground states of the Be-atom for different symmetry subspaces: artificial spin-triplet using only $s$-orbitals (left) and fully polarized, i.e.~spin-quintet (right). Increasing the dimension $d'$ of the underlying truncated setting confirms the absence of pinning since the truncation error (red error bars) is smaller than the distance $D'_{min}$ to the polytope boundary.}
\label{fig:BeS}
\end{figure}

In Appendix \ref{app:data}, the results for the natural occupation numbers are listed for five different basis set sizes ranging from $19$\ to $25$ $s$-orbitals.
Convergence on the significant digits has been achieved not only for the energy but also for the occupation numbers (see Appendix \ref{app:data}).
The results of the pinning analysis for the variational minimum of the Be atom without angular degrees of freedom are shown in Fig.~\ref{fig:BeS}.
There, we present the minimal distances $D'_{min}$ of the truncated vectors $\vec{\lambda}'\equiv(\lambda_j)_{j=1}^{d'}$ for the
reduced settings $(N',d')$ with $N'=4$, $d'=8,9,10$.
The respective truncation errors defined in Eq.~\eqref{eq:trunc} are indicated in the form of error bars.
The results strongly suggest the absence of pinning in the correct, full setting (recall Eq.~(\ref{eq:trunc})), $D_{min}>0$.
At the same time, we again find non-trivial quasipinning: $\vec{\lambda}$ is about five times closer to the polytope boundary than what one may expect from the approximate saturation of some Pauli constraints.

\begin{figure}[]
\centering
\includegraphics[width=4.0cm]{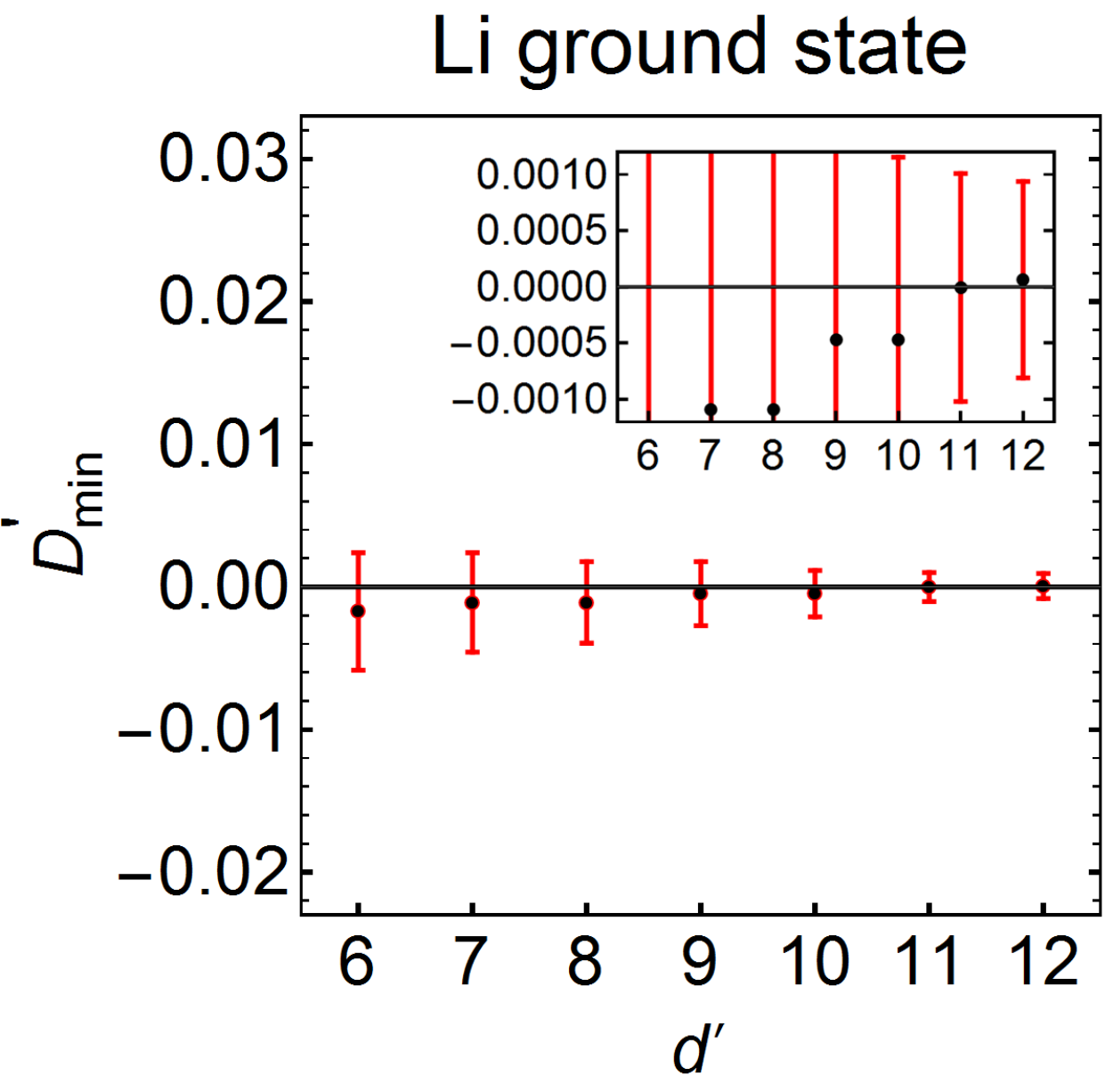}\hspace{0.3cm}
\includegraphics[width=4.0cm]{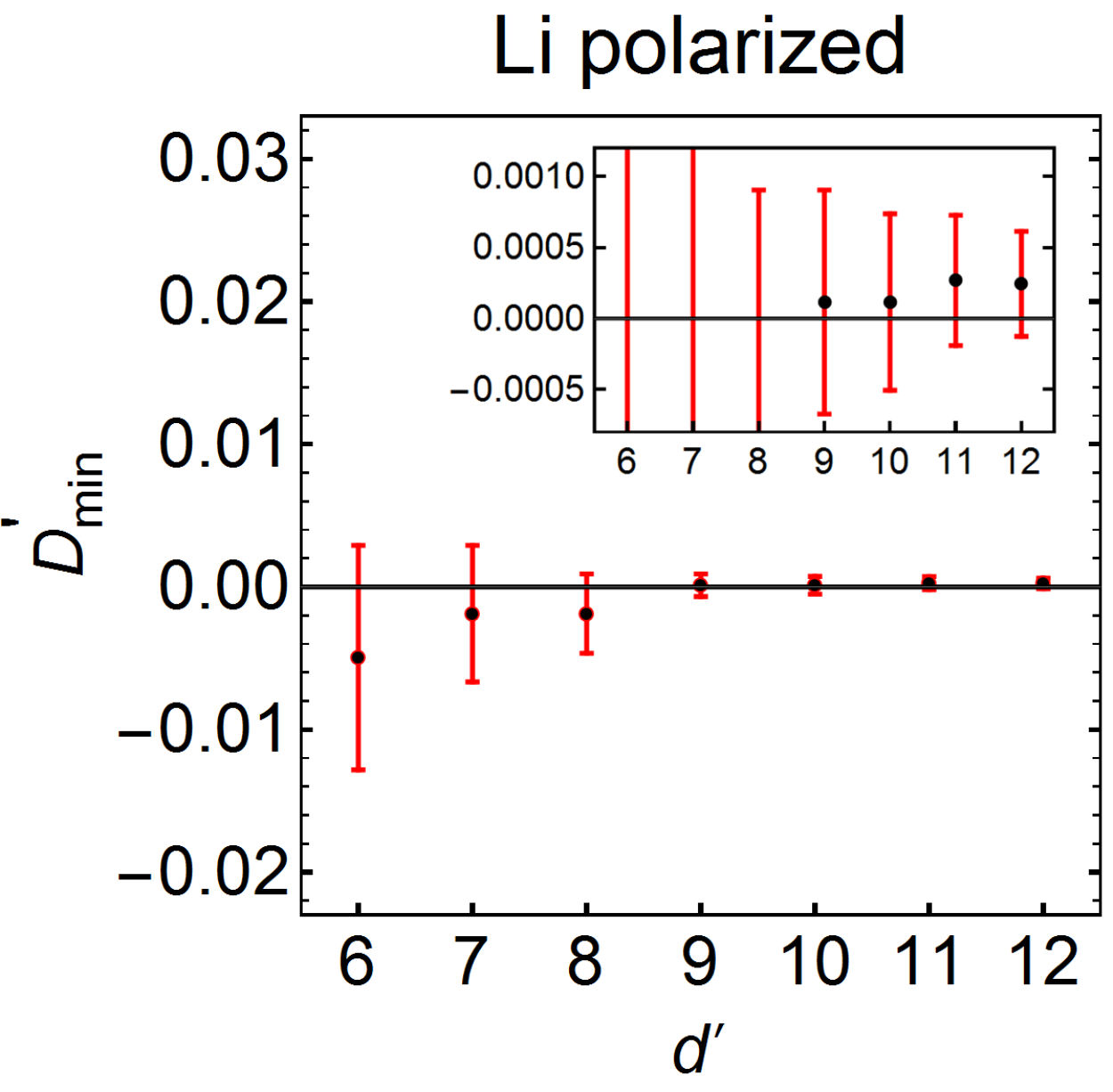}
\caption{Results of different truncated pinning analyses for the variational ground states of the Li-atom for different symmetry subspaces: spin-doublet, i.e.~overall ground state (left) and fully polarized, i.e.~spin-quadruplet (right). Presence of pinning cannot be ruled out since $D'_{min}$ is smaller than the respective truncation errors (red error bars).}
\label{fig:Li}
\end{figure}

We continue with an analysis of the doublet ground state ($S$=1/2) of the Li atom. To this end, we carried out a
non-relativistic FCI optimization with a Gaussian-type atomic orbital basis set exceeding quintuple-zeta quality, allowing
us to approach the present variational upper bound of the nonrelativistic total energy in the Li $^2S$\ ground state  of 7.4780603239101437 hartree \cite{wang11} within sub-millihartree accuracy. The numerical data and further computational details are presented in Appendices \ref{app:FCI}, \ref{app:data}.
We then performed a truncated pinning analysis in the largest setting, $(N',d')=(3,12)$ for which the GPCs are known. To reveal a qualitative trend for $D'_{min}$ for larger $d'$ and in particular to extrapolate to the full setting $(3,962)$, we also present the results for the smaller truncated settings with $d'=6-11$ in Fig.~\ref{fig:Li}: First, for $d'=6-11$ we see that the truncated vectors $\vec{\lambda}'\equiv(\lambda_j)_{j=1}^{d'}$ lie outside the respective polytopes for the settings $(3,d')$, indicated by the negative signs of $D'_{min}$. As a consistency check, we also observe that the truncation error (cf.~Eq.~\eqref{eq:trunc}) still allows for a positive distance $D_{min}$ in the full setting as required since $\vec{\lambda}\equiv (\lambda_j)_{j=1}^{962}$ lies inside the polytope of the full setting.
Second, in the largest possible truncated setting $(3,12)$ we find a minimal distance of $D'_{min}=6.46\cdot 10^{-5}$.
Since the truncation error $\varepsilon'=8.73\cdot 10^{-4}$ is much larger than $D'_{min}$, exact pinning in the full setting cannot be ruled out.
We can conclude that $\vec{\lambda}$ has a distance $D_{min}$ to the polytope boundary in the full setting $(3,962)$ of \emph{at most} $9.38\cdot 10^{-4}$ following from (\ref{eq:trunc}).
Third, in the setting $(3,12)$, we find that $\vec{\lambda}$ lies closer by a factor of $60$ to the polytope boundary than to the boundary of the surrounding Pauli simplex.

The case of a total spin $S$=3/2 is treated in the same fashion. The pinning analysis in the largest possible truncated setting leads to a minimal distance $D'_{min}=2.44\cdot10^{-4}$ with a truncation error $\varepsilon' =3.70\cdot 10^{-4}$. The presence of pinning in the full setting can again not be ruled out given the truncation error (cf.~Eq.~\eqref{eq:trunc}). Yet, the trend for $D'_{min}$ for $d'\leq 12$ shown in Fig.~\ref{fig:Li} and the fact that the worst-case truncation error for $d'=12$ is only slightly larger than the respective $D'_{min}$ provides plausible evidence that none of the GPCs in the full setting are saturated, i.e.~$D_{min}>0$. The analysis of the non-triviality of these findings shows that the GPCs have a physical relevance since their approximate saturation is stronger by a factor $37.7$ than the one of the Pauli constraints.

We complete our investigation of the role of the GPCs in small atoms by returning to the Be atom. There are three spin sectors, corresponding to the total spins $S=2,1,0$. The case of a singlet state, $S=0$, does not allow for any non-trivial relevance of the GPCs. This is based on the fact that for singlet states and in general for states of evenly many fermions with doubly degenerate natural occupation numbers, the GPCs coincide with the original Pauli constraints \cite{Smith}. From a geometric viewpoint, the polytope $\mathcal{P}$ and the larger Pauli simplex $\Delta$ coincide after their restriction to the hyperplane described by $\lambda_1=\lambda_2,\, \lambda_3=\lambda_4,\ldots$. Since we already analyzed the lowest-lying $S$=1 state of Be, we now consider the case of the lowest-lying quintet state ($S$=2). The numerically exact natural occupation numbers for fully polarized electrons are listed in Appendix \ref{app:data} and the results of the truncated pinning analysis are illustrated in Fig.~\ref{fig:BeS}. Since $1-\lambda_1=3.34\cdot10^{-4}$ is (much) smaller than $\lambda_{11}=\lambda_{12}=2.14\cdot10^{-3}$ we choose as truncated settings $(3,d')$ with $d'=6-12$ rather than $(4,d')$ with $d'\leq 10$. The respective truncation errors are given again by Eq.~\eqref{eq:trunc} but now with $N-N'=1>0$. Since for the largest truncated setting, $(N',d')\equiv(3,12)$, the truncation error $\varepsilon' = 1.74\cdot 10^{-3}$ is smaller than the minimal distance $D'_{min}=4.56\cdot 10^{-3}$ found in that setting,
no pinning is present in the full setting $(4,800)$, i.e.~$D_{min}>0$.
Moreover, one can show that this weaker quasipinning is completely trivial: This follows immediately from the fact that the approximate saturation of some Pauli constraints is comparable.

\section{Acknowledgements}
We are grateful to T.\hspace{0.5mm}Farrow, D.\hspace{0.5mm}Jaksch, A.\hspace{0.5mm}Klyachko, M.\hspace{0.5mm}Nakata, M.\hspace{.5mm}Walter for helpful discussions. MA and CS thank A.\hspace{0.5mm}Klyachko for the hospitality in Ankara in Nov 2015. The numerical calculations were performed on the Brutus and Euler clusters at ETH Zurich and the programs \verb"LiE" \cite{LiE} and \verb"Convex" \cite{Convex} were essential for the calculation of the GPCs of the larger settings. We acknowledge financial support from the Swiss National Science Foundation (Grant P2EZP2 152190), the Oxford Martin Programme on Bio-Inspired Quantum Technologies, the UK Engineering and Physical Sciences Research Council (Grant EP/P007155/1) (CS),
the European Research Council (ERC Grant Agreement no 337603), the Danish Council for Independent Research (Sapere Aude) and VILLUM FONDEN via the QMATH Centre of Excellence (Grant No. 10059) (MC), the Excellence Initiative of the German Federal and State Governments (Grants ZUK 43, ZUK 81), the DFG through CRC 183 (project B01) (DG) and the Swiss National Science Foundation (Grant 20020\_169120) (MR).

\bibliography{refs4}

\begin{appendix}

\section{Quantifying quasipinning and concept of truncation}\label{app:truncation}
As already explained in the main text, the GPC together with the ordering constraints $\lambda_1\geq \ldots\geq \lambda_d\geq 0$ and the normalization $\sum_{j=1}^d\lambda_j=N$ give rise to a polytope $\mathcal{P}^{(N,d)}\subset \R^d$. In principle, we may quantify the strength of quasipinning of $\vec{\lambda}\in \mathcal{P}^{(N,d)}$ by the $l_1$-distance of $\vec{\lambda}$ to the polytope boundary. Yet, a few comments are in order. First, by ``boundary'' we refer in this context only to that part of the total boundary of the polytope which corresponds to saturation of some GPC (but not to saturation of an ordering constraint $\lambda_i-\lambda_{i+1}\geq 0$). Second, it has proven convenient to determine for each GPC $D$ the minimal distance \emph{not} to the \emph{polytope facet} $F_D$ (corresponding to $D\equiv 0$) but to the \emph{hyperplane} $E_D$ obtained by linearly extending $F_D$ and after relaxing the normalization (allowing for a minimum which is attained outside of $\mathcal{P}^{(N,d)}$). One verifies then \cite{CSQuasipinning} that the $l_1$-distance of $\vec{\lambda}$ to that hyperplane $E_D$ follows as
\begin{equation}\label{eq:l1}
\mbox{dist}_1(\vec{\lambda},E_D)= \frac{D(\vec{\lambda})}{\mbox{max}_{1\leq j\leq d}\left(|\kappa^{(j)}|\right)}\,.
\end{equation}

For (quasi)pinning analyses in practice one faces a major problem. On the one hand, the complete family of GPC is known so far only for the settings $(N,d)$ up to $d=10$ \cite{Altun} and also, as calculated for this work, for $(3,11),(3,12)$. On the other hand, most few-fermion models are based on very large or even infinite-dimensional one-particle Hilbert space. A common (see, e.g., Refs.~\cite{Kly1,BenavLiQuasi,Mazz14,BenavQuasi2,RDMFT}), but less optimal way to circumvent that problem is to truncate the one-particle Hilbert space $\mathcal{H}_1$ from the very beginning to just $d\leq10$ dimensions and restrict the Hamiltonian to the corresponding subspace $\wedge^{N}[\mathcal{H}_1^{(d)}]$. For most physical models, however, this drastic approximation does not allow one to conclusively explore the occurrence of (quasi)pinning for the exact ground state as our work has shown. Besides the objection that the system may be in general too correlated in order to justify such a truncation also an erroneous choice of $\mathcal{H}_1^{(d)}$ can lead to dubious results on (quasi)pinning.

A systematic way to avoid the error related to the choice of $\mathcal{H}_1^{(d)}$ is to implement such a truncation to a small $d$ \emph{after} having obtained a sufficiently accurate approximation for the exact ground state (as it has been done in our work). The general structure underlying the concept of truncation is then the following:
\begin{equation}\label{eq:Prelation}
\mathcal{P}^{(N,d)}|_{
\tiny \begin{array}{l}
\lambda_1=\ldots =\lambda_r=1\\
\lambda_{d+1-s}=\ldots =\lambda_{d}=0
\end{array}
} = \,\mathcal{P}_{\rm emb}^{(N',d')} \equiv \,  \vec{1}_r \oplus \mathcal{P}^{(N',d')} \oplus \vec{0}_s \,,
\end{equation}
where $0\leq r \leq N$, $0\leq s \leq d-N$, $N'\equiv N-r$ and $d' \equiv d-r-s$. On the right side of Eq.~(\ref{eq:Prelation}), the polytope
$\mathcal{P}^{(N',d')}$ is embedded into $\R^d$ by adding as first $r$ coordinates $1$'s (denoted by $\vec{1}_r$) and as the $s$ last coordinates $0$'s (denoted by $\vec{0}_s$). Relation (\ref{eq:Prelation}) states that restricting the polytope $\mathcal{P}^{(N,d)}$ to the hyperplane defined by $\lambda_1=\ldots =\lambda_r=1$, $\lambda_{d-s+1}=\ldots =\lambda_{d}=0$ leads essentially (i.e.~up to the embedding) to the corresponding polytope of the setting $(N-r,d-r-s)\equiv (N',d')$. In particular, it implies for every facet $F'$ of the polytope $\mathcal{P}^{(N',d')}$ (corresponding to a GPC $D'$) that its embedding $F'_{\rm emb}$ into $\R^d$ according to (\ref{eq:Prelation}) is a subset of a specific facet $F$ of the full polytope $\mathcal{P}^{(N,d)}$ (corresponding to the GPC $D$). On the level of the GPC this implies a relation between $D'$ and $D$ of the form \cite{CS2016a}
\begin{align}\label{eq:relGPC}
D(\vec{\lambda})&= \kappa^{(0)}+\sum_{j=1}^d \kappa^{(j)}\lambda_j=\sum_{j=1}^r \kappa^{(j)}(\lambda_j-1)\\
& \quad  +D'(\vec{\lambda}')+\sum_{k=d-s+1}^d\kappa^{(k)}\lambda_k\,. \nonumber
\end{align}
Hence, whenever the vector $\vec{\lambda}_{\rm emb}'$,
\begin{equation}
\left(\vec{\lambda}_{\rm emb}'\right)_j\equiv
     \begin{cases}
    0,& \text{if } j> d-s\\
    \lambda_j,& \text{if } r< j\leq d-s\\
    1, & \text{if } j\leq r
\end{cases}
\end{equation}
has a small $l_1$-distance to $F'_{\rm emb}$ and the ``truncation error''
\begin{equation}\label{eq:truncer}
\varepsilon'\equiv \mbox{dist}_1(\vec{\lambda}'_{\rm emb},\vec{\lambda})=\sum_{j=1}^r(1-\lambda_j)+\sum_{k=d-s+1}^{d}\lambda_k
\end{equation}
is small, the full vector $\vec{\lambda}$ has also a small $l_1$-distance to the facet $F$. In particular, by using the triangle inequality for the $l_1$-norm and by introducing the truncated vector $\vec{\lambda}'\equiv (\lambda_j)_{j=r+1}^{d-s}$, we find for the distances to the linear extensions ($E,E'_{\rm emb}, E'$) of the respective facets ($F,F'_{\rm emb}, F'$)
\begin{eqnarray}\label{eq:est1}
\mbox{dist}_1(\vec{\lambda},E) &\leq & \mbox{dist}_1(\vec{\lambda}'_{\rm emb},E) +  \mbox{dist}_1(\vec{\lambda}'_{\rm emb},\vec{\lambda}) \nonumber \\
&\leq & \mbox{dist}_1(\vec{\lambda}'_{\rm emb},E'_{\rm emb}) +  \varepsilon' \nonumber \\
&=& \mbox{dist}_1(\vec{\lambda}',E') +  \varepsilon' \,.
\end{eqnarray}
In the second line, we have used $E'_{\rm emb} \subset E$ and the definition of the truncation error $\varepsilon'$ (\ref{eq:truncer}). Conversely, by using Eqs.~(\ref{eq:l1}), (\ref{eq:relGPC}) one finds
\begin{eqnarray}\label{eq:est2}
\mbox{dist}_1 &(\vec{\lambda},E)&= \frac{D(\vec{\lambda})}{\max_{1\leq j\leq d}\left(|\kappa^{(j)}|\right)}\nonumber \\
&=& \frac{D'(\vec{\lambda}')}{\max_{1\leq j\leq d}\left(|\kappa^{(j)}|\right)}\\
&&+\frac{\sum_{j=1}^r \kappa^{(j)}(\lambda_j-1) +\sum_{k=d-s+1}^d\kappa^{(k)}\lambda_k}{\max_{1\leq j\leq d}\left(|\kappa^{(j)}|\right)}\nonumber \\
&=& \frac{\max_{r+1\leq j \leq d-s}\left(|\kappa^{(j)}|\right)}{\max_{1\leq j\leq d}\left(|\kappa^{(j)}|\right)}\frac{D'(\vec{\lambda}')}{\max_{r+1\leq j\leq d-s}\left(|\kappa^{(j)}|\right)} \nonumber\\
&&+\frac{\sum_{j=1}^r \kappa^{(j)}(\lambda_j-1) +\sum_{k=d-s+1}^d\kappa^{(k)}\lambda_k}{max_{1\leq j\leq d}\left(|\kappa^{(j)}|\right)} \\
&\geq& \frac{\max_{r+1\leq j \leq d-s}\left(|\kappa^{(j)}|\right)}{\max_{1\leq j\leq d}\left(|\kappa^{(j)}|\right)}\cdot\mbox{dist}_1(\vec{\lambda}',E')
-  \varepsilon' \nonumber \\
&\equiv& c\cdot\mbox{dist}_1(\vec{\lambda}',E')-  \varepsilon' \,.
\end{eqnarray}
Clearly, one finds for the geometric prefactor $c \leq 1$. By comparing various settings for which the GPC are already known, it seems to be plausible to assume $c\equiv1$ for the case of $d'$ sufficiently large. In that case, one can combine estimates (\ref{eq:est1}), (\ref{eq:est2}) to
\begin{equation}\label{eq:rell1}
\left|\mbox{dist}_1(\vec{\lambda},E)-\mbox{dist}_1(\vec{\lambda}',E')\right| \leq \varepsilon'\,.
\end{equation}
Comparing the minimal distances
$D_{min}=\min_{E}\left[\mbox{dist}_1(\vec{\lambda},E)\right]$ and $D'_{min}=\min_{E'}\left[\mbox{dist}_1(\vec{\lambda}',E')\right]$
we minimize both estimates (\ref{eq:est1}) and (\ref{eq:est2}) with respect to all
$E'$ (i.e.~all GPC $D'$ of $(N',d')$) and for each $E'$ with respect to all $E$ (i.e.~GPC $D$ of $(N,d)$)
containing $E'_{\rm emb}$. This eventually leads to 
\begin{equation}\label{eq:2}
\left|D_{min}-D'_{min}\right| \leq \varepsilon'\,.
\end{equation}
The concept of truncation follows from estimate (\ref{eq:2}): Natural occupation numbers sufficiently close to $0$ and $1$ can be discarded and (quasi)pinning of the truncated vector $\vec{\lambda}'$ can be explored in the truncated setting $(N',d')$ of the remaining natural occupation numbers.  The error of this truncated pinning analysis is then given by Eq.~(\ref{eq:truncer}) and therefore allows one to estimate the minimal distance $D_{min}$ of $\vec{\lambda}$ to the polytope boundary in the correct, full setting $(N,d)$.

\section{Analysis with the $Q$-parameter}\label{app:Q}
We briefly comment on the concept of the $Q$-parameter. Since the polytope $\mathcal{P}$ defined by the GPC is a proper subset of the Pauli simplex $\Delta$, $\mathcal{P}\subset \Delta$, pinning and quasipinning by GPC can in some cases be a consequence of (quasi)pinning by the Pauli constraints. From a geometrical viewpoint this is obvious: Whenever $\vec{\lambda}\in \mathcal{P}\subset \Delta$ is close to $\partial \Delta$ it is at least as close to the boundary $\partial \mathcal{P}$. Hence, the important question is whether (quasi)pinning by GPC is stronger than possible (quasi)pinning by the Pauli constraints, i.e.~whether the distance of $\vec{\lambda}$ to the polytope boundary is much smaller than its distance to the boundary of the surrounding Pauli simplex $\Delta$. Actually, this is even more subtle: In some cases, the strongest (quasi)pinning, let's say to the facet $F_D$ corresponding to saturation of the GPC $D$, might be trivial, yet weaker quasipinning to another polytope facet $F_{D'}$ might be non-trivial. In that case the relevance of the family of GPC would be based on GPC $D'$. Hence, the non-triviality of pinning and quasipinning by GPC needs to be addressed for every GPC $D$, i.e.~for every facet $F_D$, separately. This leads to a highly involved geometrical problem which has been solved in Ref.~\cite{CSQ}. In particular, for every GPC $D$ it has been determine how many occupation numbers need to be pinned by the Pauli constraints to the values 0 and 1 to enforce saturation of $D$. Then, by denoting the minimally required number of occupation numbers equal to 0 and 1 by $r$ and $s$, respectively, the flat geometry of the polytope implies relations of the form
\begin{equation}\label{eq:DvsS}
D(\vec{\lambda}) \leq c S_{r,s}(\vec{\lambda})\,.
\end{equation}
Here, $S_{r,s}$ denotes the collective Pauli exclusion principle constraint corresponding to the pair $(r,s)$, i.e.
\begin{equation}
S_{r,s}(\vec{\lambda})\equiv \sum_{i=1}^r(1-\lambda_i)+\sum_{j=d-s+1}^{d}\lambda_j\geq 0
\end{equation}
and $c$ is a geometric prefactor. The ratio of both sides in Eq.~(\ref{eq:DvsS}) then describes the non-triviality of possible quasipinning by the GPC $D$. We will resort in our work to a corresponding non-triviality measure based on such ratios, the so-called $Q$-parameter \cite{CSQ}.

For instance, the analysis in the setting $(4,10)$ has shown that (quasi)pinning by GPC 
 is enforced by (quasi)pinning of the collective Pauli constraint $S_{2,9}$, i.e.~a universal estimate of the form $D_{27}^{(4,10)}(\vec{\lambda})\leq c_{27}^{(4,10)}\,S_{2,9}(\vec{\lambda})$ follows with some geometric prefactor $c_{27}^{(4,10)}$ \cite{CSQ}. Comparing the values for $S_{2,9}$ and $D_{27}^{(4,10)}$ shows that the quasipinning by GPC $D_{27}^{(4,10)}$ for the Beryllium triplet state (without angular degrees of freedom) as discussed in the main text is non-trivial by a factor
\begin{equation}
10^{\,Q_{27}^{(4,10)}(\vec{\lambda})} \equiv \frac{c_{27}^{(4,10)}\,S_{2,9}(\vec{\lambda})}{D_{27}^{(4,10)}(\vec{\lambda})} =245\,,
\end{equation}
i.e.~the $Q$-parameter of the respective GPC takes the value $Q_{27}^{(4,10)}=2.39$. The analysis of the other 124 GPC shows that their quasipinning (which is weaker than the one of $D_{27}^{(4,10)}$) is non-trivial (if at all) by factors smaller than 245. Hence, the GPC are physically relevant for the Beryllium state at hand with an overall $Q$-parameter given by $Q^{(4,10)}(\vec{\lambda})\equiv\max_{i\leq 125}Q_{i}^{(4,10)}(\vec{\lambda})=2.39$.

\section{Calculation of the GPC for $(N,d)=(3,11), (3,12)$}\label{app:GPC311}
In this section we describe on a rather elementary level the general strategy used for calculating the GPC for a fixed setting $(N,d)$. For further technical and mathematical details on this highly involved procedure we refer the reader to Ref.~\cite{Kly3,Altun}.

As noted above, the GPC together with ordering constraints and normalization condition define a polytope $\mathcal{P}^{(N,d)}$, called the \emph{`moment polytope'}. In theory, for any setting $(N,d)$ the moment polytope is characterized by Theorem 2 and Theorem 9, respectively \cite{Kly3}. Both theorems suggest different algorithms for calculating the GPC defining the polytope $\mathcal{P}^{(N,d)}$ \cite{Kly3,Altun}. On the one hand, according the Theorem 2, the GPC are explicitly given by linear inequalities subject to a topological condition. This condition depends explicitly on so-called \emph{`test spectra'} and two permutations involved in the underlying mathematical problem (see \cite{Kly3,Altun}). In practice, this means to check the topological condition for all test spectra and permutations which, however, becomes computationally very expensive for not too small settings $(N,d)$. Hence, the algorithm based on Theorem 2 allows one to calculate some of the GPC but typically not all of them, and thus leads to an outer approximation $\mathcal{P}_{out}^{(N,d)}$ to $\mathcal{P}^{(N,d)}$.
Theorem 9, on the other hand, describes the moment polytope as the convex hull of very specific spectra. Those spectra can be obtained from the irreducible components of the symmetric powers up to a finite degree $M$ of an irreducible $U(d)$-representation, where $U(d)$ is the group of unitary operators of a $d$-dimensional Hilbert space. The full polytope $\mathcal{P}^{(N,d)}$ can be obtained in that way only if $M$ is chosen large enough. Since in practice for larger settings $(N,d)$ not all of the required irreducible components can be calculated this procedure leads to a convex hull of spectra defining a subset $\mathcal{P}_{in}^{(N,d)}$ of the full polytope $\mathcal{P}^{(N,d)}$.

Although the algorithms based on Theorem 2 and Theorem 9 are both typically too expensive for deriving \emph{all} GPC, combining them allows one to calculate all GPC for much larger settings. The following procedure has proven to be the most efficient one:
\begin{itemize}
\item[(1)] Calculate an inner polytope $\mathcal{P}_{in}^{(N,d)}$.
\item[(2)] Identify all facets of the inner polytope found in step (1) which fit into the form of Theorem 2 in \cite{Kly3}. These facets define an outer polytope $\mathcal{P}_{out}^{(N,d)}$.
\item[(3)] If both approximations coincide, $\mathcal{P}_{in}^{(N,d)}=\mathcal{P}_{out}^{(N,d)}$, one has obtained the full moment polytope $\mathcal{P}^{(N,d)}=\mathcal{P}_{in}^{(N,d)}=\mathcal{P}_{out}^{(N,d)}$. Otherwise, one needs to continue the process by calculating larger inner polytopes $\mathcal{P}_{in}^{(N,d)}$.
\end{itemize}
Further details can be found in \cite{Kly3,Altun}.

For small settings as, e.g., $(N,d)=(3,6),(3,7)$ the algorithm works very well. For larger settings, however, one needs an additional tool to calculate the full set of GPC. The reason for this is that the inner polytope $\mathcal{P}_{in}^{(N,d)}$ may have `bad facets', i.e.~there are some spectral inequalities which do not fit into the form of Theorem 2 and thus it is not possible to verify the validity of a respective topological condition. Therefore, one needs to verify numerically whether such a spectral inequality represents a proper GPC: By minimizing a linear form of the one-particle reduced density matrix one can determine in a brute-force approach extremal points (vertices) of the polytope $\mathcal{P}^{(N,d)}$ (since any linear form attains its minimum at extremal points (vertices) of the convex polytope $\mathcal{P}^{(N,d)}$). By comparing the vertices of the bad facet with those obtained by this numerical procedure, one can verify whether the respective spectral inequality represents a proper GPC or not.

A second computational problem concerns the verification of the topological condition mentioned above. The computational time for this verification depends exponentially on the length of the underlying permutations. While for the settings $(3,d)$ with $d\le11$ all required topological conditions could be verified on a personal computer, there have been 15 inequalities for the setting $(3,12)$ where even high-performance computing hardware that is accessible to us at ETH could not provide the resources necessary for the verification. Instead, for those 15 inequalities we have successfully used the numerical procedure.

Finally, we also would like to stress that the software packages \verb"LiE" \cite{LiE} and \verb"Convex" \cite{Convex} were essential for the calculation for the GPC for the settings $(3,11),(3,12)$.

\section{Variational method}\label{app:FCI}
If not indicated otherwise, all calculations for the Li and Be atoms have been carried out with the quantum-chemical program package DALTON
\cite{DALTON2015,dalt14}\ using the parallel version of the full configuration interaction (CI) module LUCITA \cite{knecht2008}.
Solving for the lowest state in a given spin and spatial symmetry, the variational FCI calculations were performed based
on a Davidson subspace approach with a residual threshold of 10$^{-7}$ hartree\ with respect to the energy.
The natural occupation numbers were subsequently obtained as eigenvalues of the one-particle reduced density matrix that was
calculated by means of the respective, converged FCI vector.

\section{Numerical data}\label{app:data}
We present in the following for each FCI calculation the largest 17 natural occupation numbers. For the Be atom restricted to the triplet spin sector with fixed magnetization $M=+1$ we chose as atomic orbital (AO) basis sets $19$ (``Be triplet 1''),  $21$ (``Be triplet 2''), $23$ (``Be triplet 3''), $24$ (``Be triplet 4'')\ and $25$ (``Be triplet 5'') $s$-type functions (angular momentum functions with $l=0$) by starting from the triply-augmented aug-cc-pVQZ basis set \cite{dunn89} and systematically adding diffuse $s$-functions while simultaneously discarding all exponents for
angular momentum functions with $l > 0$.
For the case of the lowest-lying quintet state of the Be atom we chose the aug-cc-CVQZ basis set \cite{dunn89} in uncontracted form and augmented with angular momentum functions of up to $i$-type ($l=6$), denoted as $[22s18p14d8f5g4h1i]$, comprising in total 427 primitive AO functions.
To determine the ground state of the Li atom we proceeded in two ways. First, the aug-cc-CVQZ atomic orbital basis set \cite{dunn89} in uncontracted form was augmented by adding two diffuse functions
for each $l$\ quantum number up to $l=4$ ($g$-functions) resulting in a total of 222 primitive AO functions (``Li doublet 1'', [18s12p8d6f4g]).
In a second step, we augmented the original aug-cc-pCVQZ basis set by adding
tight functions for each $l$\ quantum number up $l=4$ and further included functions of $h$- and $i$-type leading to a total of
362 (``Li doublet 2'') and 517 (``Li doublet 3'') primitive AOs that can be best summarized as [20s17p11d8f5g2h1i] and [22s20p15d10f7g4h2i] AO basis sets, respectively. Finally, the variational FCI optimization of the lowest-lying quadruplet spin state of Li (``Li quadruplet'')
was carried out by employing  the ``Li doublet 3'' basis set.

Tables \ref{tabs:basis-sets-be}\ and \ref{tabs:basis-sets-li} compile
the coefficients in the exponents of all atomic-orbital Gaussian basis sets employed in this work and described above.
The basis set for the Be triplet calculation with five s functions (see Table \ref{tab} in the main text)
is available upon request from the authors.

By comparing the results of the different Be triplet state calculations shown in Table \ref{tabs:be}\ we conclude that the natural occupation numbers have converged on at least seven digits. For the Li doublet ground and quadruplet excited states summarized in Table \ref{tabs:li}\
we observe convergence on the fourth, probably also on the fifth digit.
Our best variational energy for the Be (Li) atom in the $^1S$ ($^2S$) ground state obtained with the largest uncontracted AO basis set outlined above is -14.6667932644 hartree (-7.4778376184 hartree) which is well below millihartree accuracy compared
to the present variational upper bounds for Be \cite{puch13} and Li \cite{wang11}, respectively.

\onecolumngrid

\begin{table}[h!]
\caption{Exponents of the AO basis sets for the Be atom.
Basis sets labels according to the main text.}
\label{tabs:basis-sets-be}
\vspace{0.2cm}
\tiny{
\begin{tabular}{l|rrrrrr}\hline\hline
 & \multicolumn{6}{c}{basis set label} \\\hline
angular momentum function & \multicolumn{1}{c}{Be triplet 1} & \multicolumn{1}{c}{Be triplet
2} & \multicolumn{1}{c}{Be triplet 3} &\multicolumn{1}{c}{Be triplet 4} &\multicolumn{1}{c}{Be triplet 5} &
\multicolumn{1}{c}{Be quintet}\\
s
&   14630.0000000000  &   14630.0000000000 &   91437.5000000000  &   91437.5000000000  &   91437.5000000000 &   91437.500000\\
&    2191.0000000000  &    5477.5000000000 &   36575.0000000000  &   36575.0000000000  &   36575.0000000000 &   36575.000000\\
&     498.2000000000  &    2191.0000000000 &   14630.0000000000  &   14630.0000000000  &   14630.0000000000 &   14630.000000\\
&     140.9000000000  &    1245.5000000000 &    5477.5000000000  &    5477.5000000000  &    5477.5000000000 &    2191.000000\\
&     114.6500000000  &     498.2000000000 &    2191.0000000000  &    2191.0000000000  &    2191.0000000000 &    1245.500000\\
&      45.8600000000  &     140.9000000000 &    1245.5000000000  &    1245.5000000000  &    1245.5000000000 &     498.200000\\
&      21.7260000000  &     114.6500000000 &     498.2000000000  &     498.2000000000  &     716.5625000000 &     140.900000\\
&      16.4700000000  &      45.8600000000 &     140.9000000000  &     140.9000000000  &     498.2000000000 &     114.650000\\
&       7.8660000000  &      21.7260000000 &     114.6500000000  &     114.6500000000  &     286.6250000000 &      45.860000\\
&       6.3190000000  &      16.4700000000 &      45.8600000000  &      45.8600000000  &     140.9000000000 &      21.726000\\
&       2.8480000000  &       7.8660000000 &      21.7260000000  &      21.7260000000  &     114.6500000000 &      16.470000\\
&       2.5350000000  &       6.3190000000 &      16.4700000000  &      16.4700000000  &      45.8600000000 &       7.866000\\
&       1.0350000000  &       2.8480000000 &       7.8660000000  &       7.8660000000  &      21.7260000000 &       6.319000\\
&       0.2528000000  &       2.5350000000 &       6.3190000000  &       6.3190000000  &      16.4700000000 &       2.848000\\
&       0.1052000000  &       1.0350000000 &       2.8480000000  &       2.8480000000  &       7.8660000000 &       2.535000\\
&       0.0426100000  &       0.2528000000 &       2.5350000000  &       2.5350000000  &       6.3190000000 &       1.035000\\
&       0.0143900000  &       0.1052000000 &       1.0350000000  &       1.0350000000  &       2.8480000000 &       0.632000\\
&       0.0048597066  &       0.0426100000 &       0.2528000000  &       0.2528000000  &       2.5350000000 &       0.252800\\
&       0.0016411917  &       0.0143900000 &       0.1052000000  &       0.1052000000  &       1.0350000000 &       0.105200\\
&                     &       0.0048597066 &       0.0426100000  &      0.0426100000   &      0.2528000000  &       0.042610\\
&                     &       0.0016411917 &       0.0143900000  &      0.0143900000   &      0.1052000000  &       0.014390\\
&                     &                    &       0.0048597066  &      0.0048597066   &      0.0426100000  &       0.005756\\
&                     &                    &       0.0016411917  &      0.0016411917   &      0.0143900000  &\\
&                     &                    &                     &      0.0005542537   &      0.0048597066  &\\
&                     &                    &                     &                     &      0.0016411917  &\\\hline
p
&&&&&&6034.765625\\
&&&&&&2413.906250\\
&&&&&& 965.562500\\
&&&&&& 386.225000\\
&&&&&& 193.112500\\
&&&&&&  77.245000\\
&&&&&&  30.898000\\
&&&&&&  14.030000\\
&&&&&&  10.365000\\
&&&&&&   5.612500\\
&&&&&&   3.168000\\
&&&&&&   1.267200\\
&&&&&&   0.902400\\
&&&&&&   0.303600\\
&&&&&&   0.113000\\
&&&&&&   0.042860\\
&&&&&&   0.016250\\
&&&&&&   0.006500\\\hline
d
&&&&&&7419.43359375\\
&&&&&&2967.77343750\\
&&&&&&1187.10937500\\
&&&&&&474.84375000\\
&&&&&&189.93750000\\
&&&&&&75.97500000\\
&&&&&&30.39000000\\
&&&&&&15.23000000\\
&&&&&&6.09200000\\
&&&&&&2.43680000\\
&&&&&&1.07200000\\
&&&&&&0.44100000\\
&&&&&&0.18110000\\
&&&&&&0.05540000\\\hline
f
&&&&&&78.0125000\\
&&&&&&31.2050000\\
&&&&&&12.4820000\\
&&&&&&4.7156862\\
&&&&&&1.8862400\\
&&&&&&0.4810000\\
&&&&&&0.2550000\\
&&&&&&0.0930000\\\hline
g
&&&&&& 6.48425\\
&&&&&& 2.59370\\
&&&&&& 1.03750\\
&&&&&& 0.41500\\
&&&&&& 0.18340\\\hline
h
&&&&&& 4.7156000\\
&&&&&& 1.8862400\\
&&&&&& 0.7544960\\
&&&&&& 0.3017984\\\hline
i
&&&&&& 0.754496\\\hline
\end{tabular}
}
\end{table}
\begin{table}[h!]
\caption{Exponents of the AO basis sets for the Li atom. Basis sets labels according to the main text.}
\label{tabs:basis-sets-li}
\vspace{0.2cm}
\tiny{
\begin{tabular}{l|rrr}\hline\hline
 & \multicolumn{3}{c}{basis set label} \\\hline
angular momentum function & Li doublet 1 & Li doublet 2 & Li doublet 3\\\hline
s
&    6601.0000000000 &   41256.25000             &   41256.25000\\
&     989.7000000000 &   16502.50000             &   16502.50000\\
&     225.7000000000 &    6601.00000             &    6601.00000\\
&      64.2900000000 &    2474.25000             &    2474.25000\\
&      21.1800000000 &     989.70000             &    1410.62500\\
&       7.7240000000 &     564.25000             &     989.70000\\
&       5.6140000000 &     225.70000             &     564.25000\\
&       3.0030000000 &      64.29000             &     225.70000\\
&       1.8600000000 &      21.18000             &     160.72500\\
&       1.2120000000 &       7.72400             &      64.29000\\
&       0.6160000000 &       5.61400             &      21.18000\\
&       0.4930000000 &       3.00300             &       7.72400\\
&       0.0951500000 &       1.86000             &       5.61400\\
&       0.0479100000 &       1.21200             &       3.00300\\
&       0.0222000000 &       0.61600             &       1.86000\\
&       0.0063600000 &       0.49300             &       1.21200\\
&       0.0018220541 &       0.09515             &       0.61600\\
&       0.0005219939 &       0.04791             &       0.49300\\
&                    &       0.02220             &       0.09515\\
&                    &       0.00636             &       0.04791\\
&                    &                           &       0.02220\\
&                    &                           &       0.00636\\\hline
p
&       9.7850000000  &    5972.29003908        &   93317.03186045\\
&       6.2500000000  &    2388.91601563        &   37326.81274418\\
&       2.5930000000  &     955.56640625        &   14930.72509767\\
&       1.3700000000  &     382.22656250        &    5972.29003908\\
&       0.6870000000  &     152.89062500        &    2388.91601563\\
&       0.3672000000  &      61.15625000        &     955.56640625\\
&       0.1192000000  &      24.46250000        &     382.22656250\\
&       0.0447400000  &       9.78500000        &     152.89062500\\
&       0.0179500000  &       6.25000000        &      61.15625000\\
&       0.0075600000  &       2.59300000        &      24.46250000\\
&       0.0031840446  &       1.37000000        &       9.78500000\\
&       0.0013410238  &       0.68700000        &       6.25000000\\
&                     &       0.36720000        &       2.59300000\\
&                     &       0.11920000        &       1.37000000\\
&                     &       0.04474000        &       0.68700000\\
&                     &       0.01795000        &       0.36720000\\
&                     &       0.00756000        &       0.11920000\\
&&&       0.04474000\\
&&&       0.01795000\\
&&&       0.00756000\\\hline
d
&      10.6020000000  &     165.65625   &    1035.3515625\\
&       3.0660000000  &      66.26250   &     414.1406250\\
&       0.3440000000  &      26.50500   &     165.6562500\\
&       0.1530000000  &      10.60200   &      66.2625000\\
&       0.0680000000  &       7.66500   &      47.9062500\\
&       0.0266000000  &       3.06600   &      26.5050000\\
&       0.0104052941  &       1.22640   &      19.1625000\\
&       0.0040703062  &       0.34400   &      10.6020000\\
&                     &       0.15300   &       7.6650000\\
&                     &       0.06800   &       3.0660000\\
&                     &       0.02660   &       1.2264000\\
&                     &                 &       0.3440000\\
&&&       0.1530000\\
&&&       0.0680000\\
&&&       0.0266000\\\hline
f
&       6.6830000000&      41.76875   &     261.0546875\\
&       0.2460000000&      16.70750   &     104.4218750\\
&       0.1292000000&       6.68300   &      41.7687500\\
&       0.0552000000&       2.67320   &      16.7075000\\
&       0.0235839009&       1.06928   &       6.6830000\\
&       0.0100760939&       0.24600   &       2.6732000\\
&                   &       0.12920   &       1.0692800\\
&                   &       0.05520   &       0.2460000\\
&&&       0.1292000\\
&&&       0.0552000\\\hline
g
&       0.2380000000&       3.71875  &      23.2421875\\
&       0.1050000000&       1.48750  &       9.2968750\\
&       0.0463235294&       0.59500  &       3.7187500\\
&       0.0204368512&       0.23800  &       1.4875000\\
&                   &       0.10500  &       0.5950000\\
&&&       0.2380000\\
&&&       0.1050000\\\hline
h
&&       1.4875  &       9.296875\\
&&       0.5950  &       3.718750\\
&&&       1.487500\\
&&&       0.595000\\\hline
i
&&       1.4875  &       3.71875\\
&&&                      1.48750\\\hline
\end{tabular}
}
\end{table}

\begin{table}[h!]
\caption{Selected natural occupation numbers $\lambda$ of the Be atom in the lowest-lying triplet and quintet spin states obtained from FCI calculations with various choices of atomic orbital basis sets.}
\label{tabs:be}
\vspace{0.2cm}
$
\begin{array}{|c|c|c|c|c|c||c|}
\hline
 \text{} & \text{Be triplet 1} & \text{Be triplet 2} & \text{Be triplet 3} & \text{Be triplet 4} & \text{Be triplet 5} & \text{Be quintet} \\ \hline
 \lambda_1  & 0.999990829 & 0.999990829 & 0.999990829 & 0.999990822 & 0.999990828 & 0.999660862 \\ \hline
 \lambda_2  & 0.999985830 & 0.999985830 & 0.999985830 & 0.999985824 & 0.999985829 & 0.994201309 \\ \hline
 \lambda_3  & 0.999282259 & 0.999282260 & 0.999282260 & 0.999282259 & 0.999282253 & 0.990864390 \\ \hline
 \lambda_4  & 0.999280125 & 0.999280126 & 0.999280126 & 0.999280125 & 0.999280119 & 0.990864390 \\ \hline
 \lambda_5  & 0.000704362 & 0.000704361 & 0.000704361 & 0.000704362 & 0.000704368 & 0.004018332 \\ \hline
 \lambda_6  & 0.000698221 & 0.000698220 & 0.000698220 & 0.000698221 & 0.000698227 & 0.004018332 \\ \hline
 \lambda_7  & 0.000019918 & 0.000019918 & 0.000019918 & 0.000019919 & 0.000019918 & 0.002999224 \\ \hline
 \lambda_8  & 0.000018550 & 0.000018550 & 0.000018550 & 0.000018550 & 0.000018551 & 0.002626960 \\ \hline
 \lambda_9  & 0.000009215 & 0.000009215 & 0.000009215 & 0.000009221 & 0.000009216 & 0.002626960 \\ \hline
 \lambda_{10} & 0.000006020 & 0.000006020 & 0.000006020 & 0.000006025 & 0.000006021 & 0.002265210 \\ \hline
 \lambda_{11} & 0.000002481 & 0.000002481 & 0.000002481 & 0.000002481 & 0.000002481 & 0.002136778 \\ \hline
 \lambda_{12} & 0.000001217 & 0.000001217 & 0.000001217 & 0.000001217 & 0.000001217 & 0.002136778 \\ \hline
 \lambda_{13} & 0.000000486 & 0.000000486 & 0.000000486 & 0.000000486 & 0.000000486 & 0.000178990 \\ \hline
 \lambda_{14} & 0.000000250 & 0.000000250 & 0.000000250 & 0.000000250 & 0.000000250 & 0.000178990 \\ \hline
 \lambda_{15} & 0.000000096 & 0.000000096 & 0.000000096 & 0.000000096 & 0.000000096 & 0.000160835 \\ \hline
 \lambda_{16} & 0.000000064 & 0.000000064 & 0.000000064 & 0.000000064 & 0.000000064 & 0.000160835 \\ \hline
 \lambda_{17} & 0.000000026 & 0.000000026 & 0.000000026 & 0.000000026 & 0.000000026 & 0.000105161 \\ \hline
\end{array}
$
\end{table}
\begin{table}[h!]
\caption{Selected natural occupation numbers $\lambda$ of the Li atom in the lowest-lying doublet and quadruplet spin states obtained from FCI calculations with various choices of atomic orbital basis sets.}
\label{tabs:li}
\vspace{0.2cm}
$
\begin{array}{|c|c|c|c||c|}
\hline
 \text{} & \text{Li doublet 1} & \text{Li doublet 2} & \text{Li doublet 3} & \text{Li quadruplet} \\ \hline
 \lambda_1  & 0.999498396 & 0.999495629 & 0.999495700 & 0.999615833 \\ \hline
 \lambda_2  & 0.996626085 & 0.996629357 & 0.996632997 & 0.991665810 \\ \hline
 \lambda_3  & 0.996437236 & 0.996441281 & 0.996444995 & 0.991555324 \\ \hline
 \lambda_4  & 0.001344469 & 0.001338216 & 0.001337284 & 0.003107834 \\ \hline
 \lambda_5  & 0.001338450 & 0.001332235 & 0.001331308 & 0.003107834 \\ \hline
 \lambda_6  & 0.000637328 & 0.000633110 & 0.000631979 & 0.003080215 \\ \hline
 \lambda_7  & 0.000637328 & 0.000633110 & 0.000631979 & 0.003080215 \\ \hline
 \lambda_8  & 0.000637328 & 0.000633110 & 0.000631979 & 0.002008718 \\ \hline
 \lambda_9  & 0.000618613 & 0.000614728 & 0.000613598 & 0.001988929 \\ \hline
 \lambda_{10} & 0.000618613 & 0.000614728 & 0.000613598 & 0.000167381 \\ \hline
 \lambda_{11} & 0.000618613 & 0.000614728 & 0.000613598 & 0.000162804 \\ \hline
 \lambda_{12} & 0.000141809 & 0.000140752 & 0.000140721 & 0.000088996 \\ \hline
 \lambda_{13} & 0.000141809 & 0.000140752 & 0.000140721 & 0.000088996 \\ \hline
 \lambda_{14} & 0.000141809 & 0.000140752 & 0.000140721 & 0.000036006 \\ \hline
 \lambda_{15} & 0.000048149 & 0.000047458 & 0.000047392 & 0.000028861 \\ \hline
 \lambda_{16} & 0.000048149 & 0.000047458 & 0.000047392 & 0.000028861 \\ \hline
 \lambda_{17} & 0.000048149 & 0.000047458 & 0.000047392 & 0.000027435 \\ \hline
\end{array}
$
\end{table}
\twocolumngrid
\end{appendix}
\end{document}